%% file: desy145
\documentstyle[11pt,epsfig,twoside]{article}
\newlength{\dinwidth}
\newlength{\dinmargin}
 \newcommand{\cit}[1]{$^{#1}$}
 \newcommand{\citalt}[1]{Ref.~{#1}}
 \renewcommand{\thesection}{\arabic{section}.}
 
 \renewcommand{\theequation}{{\protect\thesection\arabic{equation}}}

 \newcommand{\mysection}[1]{
 \addtocounter{section}{1}
 \setcounter{footnote}{0}
 \setcounter{subsection}{0}
 \setcounter{equation}{0}
 \setcounter{figure}{0}
 \setcounter{table}{0}
 \bigbreak
 {\large\bf\noindent \thesection\ #1}
 \vspace{10pt} \noindent}
 
 \newcommand{\mysectionwithout}[1]{
 \bigbreak
 {\large\bf\noindent #1}
 \vspace{10pt} \noindent}

\newcommand\eqa{\begin{equation}}
\newcommand\eqb{\end{equation}}
\newcommand\Eqa{\begin{eqnarray}}
\newcommand\Eqb{\end{eqnarray}}
\newcommand{\unit}[1]{\mbox{\it #1}}              
\newcommand{\subs}[1]{\mbox{\scriptsize\it #1}}   
\newcommand{\str}{\rule{0ex}{2.7ex}}              
 
\newcommand{\eqref}[1]{Eq.~(\ref{#1})}            
\newcommand{\eqRef}[2]{Eqs.~(\ref{#1}), (\ref{#2})}

\newcommand{\fig}[1]{Figure~\ref{#1}}             
\newcommand{\tab}[1]{Table~\ref{#1}}              
 
\newcommand\Z{{\bf Z}}                            
 
\newcommand\Ac{{\cal A}}                          
\newcommand\Dc{{\cal D}}                          
 
\newcommand\U{U_{[x,\mu]}}                        
\newcommand\A{A_{[x,\mu]}}                        
 
\newcommand\dmu{\partial_\mu}                     
\newcommand\gmu{\gamma^\mu}                       
\newcommand\gf{\gamma^5}                          
 
\newcommand\Pb{\bar{P}}                           
\newcommand\xb{\bar{x}}                           
 
\newcommand\Qtop{Q_{\subs{top}} }                 
\newcommand\T{{\cal T}_2}                         
\newcommand\bb{\beta}                             

\newcommand\new{_{\subs{new}}}                    
\newcommand\old{_{\subs{old}}}                    
\newcommand\opt{_{\subs{opt}}}                    

\newcommand\ra{\begin{math}\rightarrow\ \end{math}}   

\begin{document}

\title{Topological Zero Modes in Monte Carlo Simulations}
 
\author{H. Dilger\\
        Deutsches Elektronen-Synchrotron DESY,\\
        Notkestr.\ 85, 22603 Hamburg, FR Germany}
 
\date{DESY 94-145 \\ hep-lat/9408017}

\maketitle
 {\small
    \begin{center} ABSTRACT \end{center}
  \begin{quote}
    We present an improvement of global Metropolis updating steps, the 
    instanton hits, used in a hybrid Monte Carlo simulation of the 
    two-flavor Schwinger model with staggered fermions. These hits are 
    designed to change the topological sector of the gauge field. 
    In order to match these hits to an unquenched simulation with 
    pseudofermions, the approximate zero mode structure of the lattice 
    Dirac operator has to be considered explicitly.
  \end{quote}
  \vspace{8mm}
 }

\mysectionwithout{Introduction}

The Schwinger model\cit{1} 
(SM) shows a lot of features of great interest, such 
as topological sectors of the gauge field and corresponding zero modes of the
Dirac operator according to an index theorem.\cit{2} 
Closely related to that is the anomalous breaking of the chiral 
$U(1)$-symmetry.\cit{3} 
With these properties the SM may be used as a toy model for QCD, or with a 
coupled scalar field, for the electroweak theory.\cit{4} 

For a study of the mentioned non-perturbative features in those theories 
lattice calculations with Monte Carlo (MC) methods are needed among others.
However, while topological field configurations are studied because of their 
effect on the fermionic sector, the reverse relation, i.e.\
the influence of dynamical fermions on the dynamics of topological quantities, 
is usually not treated within direct lattice calculations.
One reason is the large computational effort for systems with dynamical 
fermions close enough to the continuum limit. 

In \citalt{5} 
we studied the topological properties of the lattice SM with staggered fermions
using the hybrid Monte Carlo (HMC) algorithm with pseudofermions.\cit{6} 
As staggered fermions describe two flavors of Dirac fermions in the continuum 
limit, we compared the results with continuum calculations of the two-flavor 
SM on the two-dimensional torus $\T$.\cit{7,8} 
We found good qualitative reproduction even of the role of the topological 
sectors and zero mode contributions. 
However, we took a rather small lattice of size $6\times16$. One reason was 
that the HMC algorithm, which is often used for calculations with dynamical 
fermions, does not tunnel the potential barriers between different topological 
sectors for large $\bb$. The additional instanton hits,\cit{9} 
designed to solve this problem in pure QED$_2$, do not work on larger lattices.
 
We present here a matching of this instanton hits to dynamical fermions treated
with the pseudofermion method. In particular, whenever the topology of the gauge
field is changed, the zeromode structure of the Dirac operator changes 
simultaneously. This could be taken into account with localized instanton hits, 
which are optimized in the pseudofermion background.
The improved instanton hits overcome the bad large volume behavior. 

Before we come to this improvement, we give a brief description of the 
topological properties of the lattice SM and its implications for a MC
simulation with pseudofermions. Some useful results of \citalt{7} 
on the zero mode structure in the continuum SM are given in the appendix.
 

\mysection{Topological Properties of the Lattice Schwinger Model 
\label{topological-sec}}
 
For the lattice formulation of the SM with staggered fermions 
we start with the action ($x=(x_1,x_2), \ 0 \leq x_\mu < L_\mu,\ \ \U\in U(1)$)
\Eqa \label{action}
    S[U,\bar{\phi},\phi] = S_g + S_f
    & = & \frac{\bb}{2} \sum_{x} (2 - P_x - P_{x}^{-1})
      +   \sum_{x,y} \bar{\phi}_x \, M[U]_{xy} \, \phi_y \ ,\\
    M[U]_{xy}
    & = & \frac{1}{2} \sum_\mu \rho_{\mu,x} \left[ \U \delta_{x+e_\mu,y}
    - U_{[y,\mu]}^* \delta_{x-e_\mu,y} \right] .
\Eqb
$P_x$ is the Wilson plaquette, $\bb$ is $1/e^2$ in lattice units. For the 
fermion field we took periodic boundary conditions.%
\footnote{
It turns out that antiperiodic boundary conditions for the fermion fields can 
be absorbed in a shift of the constant part of the gauge field. Thus the choice
of fermionic boundary conditions doesn't matter in fluctuating gauge fields.
}
Since the SM is superrenormalizable, the continuum limit $a$ \ra 
$0$ is achieved for large $\bb \propto 1/a^2$. In \citalt{5} 
we found that for $\bb=10$ the continuum results are reproduced with rather 
small lattice effects. The transition to the strong coupling behavior is 
found at $\bb\simeq1$.
 
The lattice definition of the topological charge $\Qtop$ can be derived from
the lattice field strength $F_x$ by\cit{10} 
\eqa \label{Qtop}
    e^{iF_x} = P_x \ , \ F_x \in [-\pi,\pi) \ , \ \
    \Qtop = \frac{1}{2\pi} \sum_x F_x \ \in \Z \ .
\eqb
This charge has integer values and coincides with the Chern index $k$ in the 
continuum limit. Corresponding to the index theorem \eqref{index} in the 
continuum, one finds $2|\Qtop|$ approximate zero modes (AZMs) for large enough 
$\bb$.
Exact zero modes arise only in gauge field configurations of zero measure in 
configuration space.\cit{11} 
Therefore we may write down the naive formula for fermionic four-point 
functions
\eqa \label{naive}
    <\bar{\phi}_{x'} \phi_x \bar{\phi}_{y'} \phi_y> = 
    \frac{1}{Z} \int\Dc[U]\, e^{-S_g[U]} \, \det(M[U]) \label{fermionEV}
    \left[ M^{-1}_{xx'} M^{-1}_{yy'}
         -            M^{-1}_{xy'} M^{-1}_{yx'} \right] \ .
                                                                 \nonumber
\eqb
In the continuum version of such expectation values on a finite volume 
the zero modes must be treated explicitly, leading to zero mode contributions 
from the topological sectors $k=\pm1$,\cit{8} 
see also \citalt{12}
.
It was one of the aims of \citalt{5} 
to identify such contributions in lattice simulations.  
 
There are two difficulties in this task, which shouldn't be mixed up.
Firstly we expect low values of the fermion determinant $\det(M[U])$ in 
the topological sectors $\Qtop=k$, due to AZMs for large $\bb$. Therefore the 
weights $q_k$ of those sectors are small. On the other hand the AZMs should 
lead to large values of certain combinations of the fermion propagator leading 
to contributions of the sectors $\Qtop=\pm1$, which do not vanish in the 
continuum limit. In order to gain sufficient statistics in theses sectors,
it is convenient to use a modified action, in which the sectors $\Qtop\neq0$ 
are pushed. The weights $q_k$ can then be determined from the modified weights 
$q'_k$,\cit{5} 
see also \citalt{13}
.

The second difficulty arises from the potential barriers between the different 
topological sectors. The HMC algorithm did not manage to tunnel these barriers 
for large values of $\bb$. 
In \citalt{5} 
we used the so-called instanton hits,\cit{9} 
working well in pure QED$_2$. 
They are given as Metropolis updating steps with a certain global proposal. 
However, with dynamical fermions (i.e.\ with pseudofermions) the acceptance 
rate for these proposals breaks down rapidly for lattices larger than the 
$6\times 16$-lattice taken there.

In order to improve these instanton hits one has to consider the 
implications of the zero mode structure in an algorithm with pseudofermions. In
the pseudofermion method the weight factor for the gauge field configurations 
is substituted by 
\eqa \label{pfaction}
    \det(M[U]) = \int\Dc[\Phi^\dagger,\Phi]
    \ e^{-S_p[U,\Phi]} , \, 
    S_p[U,\Phi] = \Phi^\dagger(M[U]M^\dagger[U])^{-1}\Phi \, .
\eqb
Here the pseudofermion field $\Phi$ is a complex bosonic auxiliary field. For
staggered fermions it can be taken only on the even sites of the lattice, thus
avoiding the additional spectrum doubling due to the product 
$M[U]M^\dagger[U]$.\cit{14} 
So we must only consider the even AZMs of this product, i.e.\ we have 
$|\Qtop|$ even AZMs $\chi^i[U]$ for large $\bb$. 
The HMC algorithm now generates configurations $[U,\Phi]$ determined by the 
action $S=S_g+S_p$ in \eqRef{action}{pfaction}. As this action contains 
$M[U]$ to a negative power, the AZMs lead to a strong suppression of 
configurations with $\Phi$ not orthogonal on $\chi^i[U]$
\eqa \label{orthogonal}
    |< \Phi, \chi^i[U] >| \simeq 0 \ .
\eqb


\mysection{Instanton Hits with Pseudofermions}

The instanton hits proposed in \citalt{9} 
are metropolis update steps of the gauge field with a special global proposal,
designed to change the topological charge $\Qtop$ by $\pm1$. Detailed balance 
of such Metropolis steps can be most easily attained by a symmetric proposal 
probability $P_P$
\eqa \label{PP}
    P_P(U,\Phi \rightarrow U',\Phi') = P_P(U',\Phi' \rightarrow U,\Phi)
\eqb
and an acceptance probability for these proposals in the Metropolis decision 
\eqa \label{PA}
    P_A(U,\Phi \rightarrow U',\Phi') = \unit{min}(1,e^{-\Delta S}) \, , \ 
    \Delta S = S_g[U'] + S_p[U',\Phi'] - S_g[U'] - S_p[U,\Phi] \ .      
\eqb
The proposal of $[U',\Phi']$ is implemented by an interpolating gauge field
$\Delta U$ with $\Qtop=1$
\eqa
    U'_{[x,\mu]} = \Delta\U \cdot \U \ , \ \ \Phi'_x = \Phi_x \ .
\eqb
The proposal probability $P_P$ is symmetric, if $\Delta U$ does not depend on
$U$, and if the inverse change $(\Delta U)^{-1}$ is proposed with equal 
probability. 
The computational cost of these instanton hits is essentially given by the 
determination of $S[U',\Phi']$ for the Metropolis decision \eqref{PA}, 
including a Conjugate Gradient (CG) solution of the linear equation 
$(M[U]M^\dagger[U])\,X=\Phi$. The success of these hits is limited by the 
mean acceptance probability $\Pb_A$.

In pure QED$_2$ the interpolating gauge fields are determined by an 
optimization of $P_A$. Since we want to choose $\Delta U$ independent of $U$, 
we assume the most likely starting configuration $U$, i.e.\ the minimum of the 
old sector $\Qtop=n$. The aim is to propose $U'$ such that the action is 
minimal in the sector $\Qtop=n\pm1$. Thus $\Delta U$ interpolates between two 
neighboring minima, it is an (euclidean) instanton. The minima are given by a 
constant field strength 
\eqa
    F^{\subs{min}}_x \ = \ 2\pi\Qtop/(L_1 L_2) \ , \mbox{ so } 
    \Delta F_x \ = \ \pm 2\pi/(L_1 L_2) \ .
\eqb
 
With dynamical fermions the situation changes completely. Since the 
pseudofermions are held fix, the gauge of the interpolating field $\Delta U$ 
is no longer arbitrary. The choice of a minimal gauge for $\Delta U$ is not 
unique, because the necessary kink in a topologically non-trivial configuration
can be situated everywhere on the lattice, see e.g.\ \citalt{15}
.
As well, $S_p[U,\Phi]$ depends strongly on the constant part of $U$, the toron
field $t_\mu$, see \eqref{Hodge}. In \citalt{5} 
it was therefore necessary to combine the instanton hits with simultaneous 
toron shifts 
\eqa
    \U \ \rightarrow \ e^{i\Delta t_\mu} \, \U \ , \ \ 
    \Delta t_\mu = R\, 2\pi/L_\mu \ , 
\eqb
where $R$ is a random number between $-1/2$ and $1/2$. 
However, also with this slight modification, the acceptance rate $P_A$ 
breaks down rapidly for larger volumes.

For large enough $\bb$ and thus sharp AZMs, the main part in $\Delta S$ 
comes from the AZMs in the new gauge field configuration $\chi^i\new[U']$. 
They are no longer orthogonal on the pseudofermion field $\Phi$, leading to 
a high $S_p[U']$ and thus to low acceptance rates $P_A$ in \eqref{PA}.
In order to control this effect we measured the acceptance rate $P_A$, and the
reduced acceptance rate $P_R$ for proposals $\Qtop=0 \rightarrow \Qtop=\pm1$. 
In the definition of the reduced acceptance rate, the effect of the lowest 
mode is separated
\eqa \label{reduced}
    P_R \equiv \unit{min}(1,e^{-\Delta S_R[U,\Phi]}) \ , \ \ 
    S_R[U,\Phi] = S[U,\Phi] \ - \ |<\Phi,\chi_0>|^2 \ \lambda_0^{-1} \ , 
\eqb
where $\chi_0$ is the lowest normalized even eigenmode of $M[U]M^\dagger[U]$, 
with eigenvalue $\lambda_0$. 
It appears that the strongest suppression of the acceptance rate is in fact due
to the zero mode part of $\Delta S_p$, see \tab{acceptance-tab}. 
\begin{table}
\begin{center}
\caption{ 
         Mean acceptance rate $\Pb_A$ and mean reduced acceptance rate $\Pb_R$ 
         for $\Qtop=0$ $\rightarrow$ $\pm1$ proposals on a 
         $12\times12$-lattice, $\bb=4$.
\label{acceptance-tab}}
\vspace{3mm}
\begin{tabular}{c|c|c}
  \hline
    & old (global) instanton hits & local instanton hits ($r=6$) with 
                                                         $\xb$-opt.\ \str\\
  \hline
   $\Pb_A$: & $0.0005(1)$ & $0.0043(3)$   \str\\
   $\Pb_R$: & $0.0262(8)$ & $0.1040(14)$  \str\\
  \hline
\end{tabular}
\end{center}
\end{table}
In this sense we want to achieve
\eqa
    <\Phi,\chi^i\old> \simeq 0 \ \ \rightarrow \ \ <\Phi,\chi^j\new> \simeq 0 \
\eqb
for as many as possible new AZMs $\chi^j\new$, see \eqref{orthogonal}. 
Let us therefore consider the zero mode structure in more detail. 

In the continuum SM there is a multiplicative relation 
between the even zero modes in the gauge field backgrounds $A_\mu(x)$,
$\Delta A_\mu(x)$, and $A'_\mu(x)=A_\mu(x)+\Delta A_\mu(x)$, with Chern indices
$n$$>$$0,\,1$ and $n$$+$$1$, respectively, see \eqref{multi}.
If the zero mode $\tilde{\chi}$ in $\Delta A_\mu(x)$ is essentially constant on
most of the torus
\eqa \label{constant}
    \tilde{\chi}_a^b(x) \simeq c\,\omega_a^b \mbox{ for } x \not\in \Ac \ ,
\eqb
it follows from this multiplicative structure, that $n$ of the $n+1$ zero modes
$\chi^i\new$ in $A'$ are approximately given by the old zero modes $\chi^i\old$
everywhere, but in the region $\Ac$. 
With \eqref{multi} we can estimate
\eqa \label{estimate}
    < \Phi, \chi^i\new >  \ \leq  \ c < \Phi, \chi^i\old > \ + \
    m < |\Phi| , |(\tilde{\chi}-c\,\omega)| > \ , 
\eqb
where $m$ is an upper bound for $|\chi\old(x)|$. On the lattice, the first part
of the rhs of \eqref{estimate} vanishes approximately in the simulation, see
\eqref{orthogonal}. The second part is small if the region of strong deviations
$\Ac$ in \eqref{constant} is small.
 
Thus, on the lattice we look for interpolating fields with an AZM showing the 
behavior of \eqref{constant}. The region of strong deviations $\Ac$ should be 
as small as possible. We take local instanton configurations with link 
variables $\Delta \U^r = \exp(i\Delta\A^r)$ 
\eqa \label{define}
   \Delta\A^r = \left\{ \begin{array}{cl}
                   (-\pi/r^2) \left[ \epsilon_\mu^{\ \nu} x_\nu 
                   - (-1)^{\mu} \, r \, \delta_{{x_\mu},r-1} \right]  
                &  \mbox{for } \ 0 \leq x_\mu < r  \\
                &                                 \\
              0 &  \mbox{elsewhere } \end{array} \right. \ .
\eqb
The field strength then is
$
    F_x = 2\pi/r^2 \ \mbox{ for } \ 0 \leq x_\mu < r \, , \ 
    F_x = 0 \ \mbox{ elsewhere} 
$.
\begin{figure}[htb] 
\vspace{120mm}
\begin{flushleft}
\begin{picture}( 50,100)
\put( 40.00,395.00){\makebox(0,0)[lc]{$1 - A(x)$}}
\put( 40.00,155.00){\makebox(0,0)[lc]{$\varphi(x)$}}
\epsfig{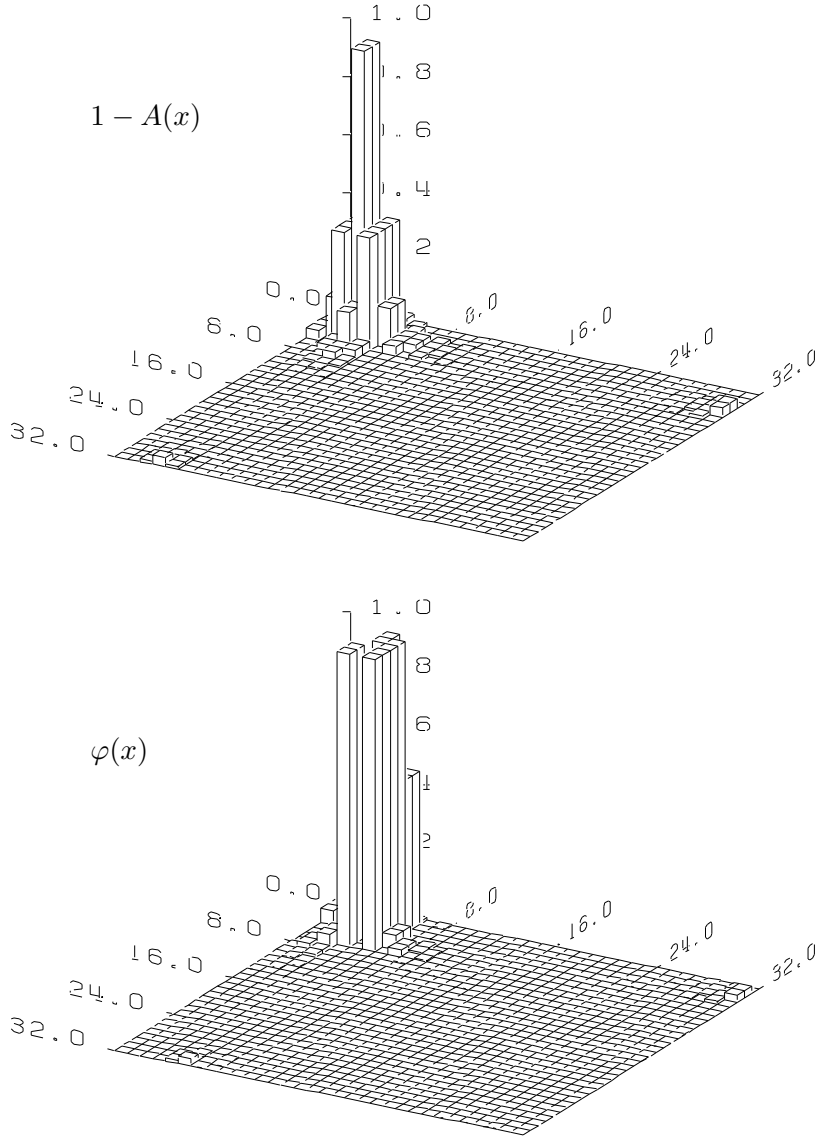}
\end{picture}
\end{flushleft}
\vspace{-3mm}
\caption{Absolute value $A(x)$ (plotted as $1-A(x)$) and phase $\varphi(x)$ of
         the AZM $\tilde{\chi}^6(x) / \Omega^6(x)$ on a $32\times32$-lattice, 
         $\Omega^6(x) = 0.0463 (-0.0424i)$ for $x_1,x_2 = \unit{even (odd)}$.
         $\chi$ is only defined on the even sites, for simplicity the values
         on odd sites are given by the neighboring even sites.   
\label{chitil-fig}}  
\end{figure}
The (even) AZM $\tilde{\chi}^r(x)$ in this configuration can be
seen numerically to be 
\eqa \label{lattice-constant}
    \tilde{\chi}^r(x) = f(x) \Omega^r(x) \ , 
\eqb 
where $f(x)\simeq 1$, if $x$ is far enough from the region
of non-vanishing gauge field $\A$, see \fig{chitil-fig}. 
$\Omega^r(x) = a^0 (a^{12})$ for $x_1,x_2 = \unit{even} \unit{(odd)}$
corresponds to the differential form $c\,\omega$ in the continuum for 
$a^{12}=-ia^0$, see \eqref{omega}. This relation gets approximated on the 
lattice for increasing $r$.
Other choices of local configurations ($\U = 1$ for $x_\mu \not\in [0,r)$) lead
to zero modes with the same principal behavior, yet with bigger deviations.

\begin{figure}[htb]
\begin{flushleft}
\vspace{-161mm}
\vbox{ \epsfig{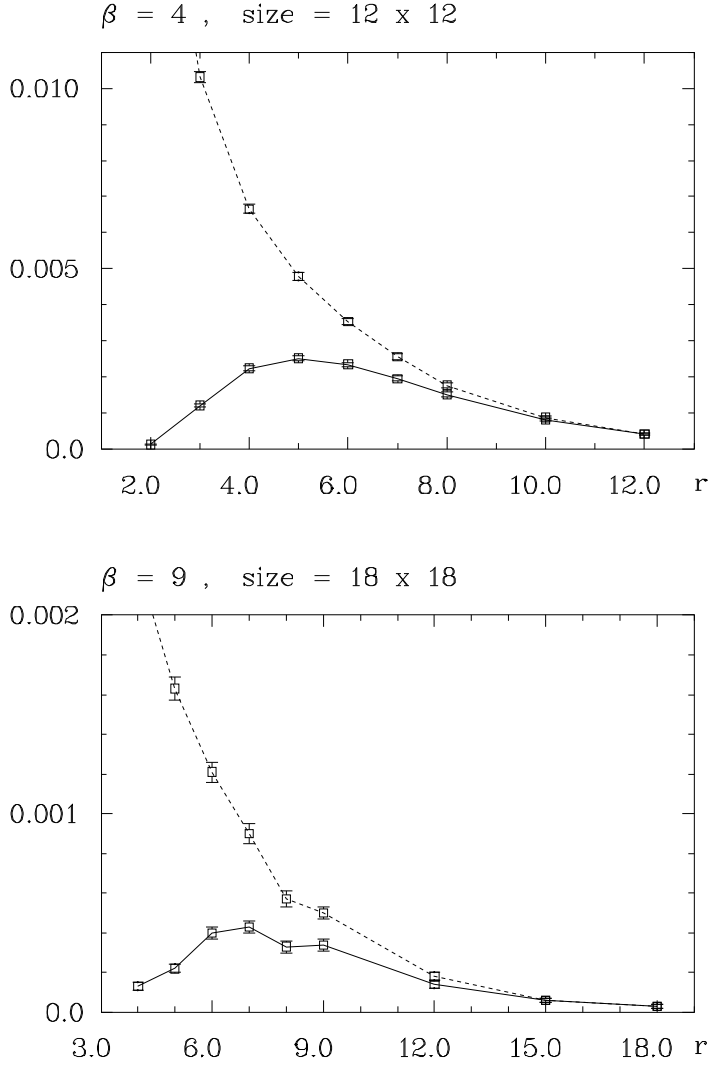} }
\end{flushleft}
\vspace{-20mm}
\caption{The mean acceptance rates $\Pb_A(r)$ (full) and $\Pb^f_A(r)$ (dashed).
\label{PfA-fig}}  
\end{figure}

Plotting the mean acceptance rate $\Pb_A(r)$ of such local instanton hits
shows the effect of the localization, see \fig{PfA-fig}. 
The change in the pseudofermion action $S_p$ favors small $r$, as indicated
by the mean fermionic acceptance rate $\Pb^f_A(r)$, $P^f_A = \unit{min}
(1,\exp(S_p-S_p')\,)$. This effect gets compensated by $S_g$ for $r<r\opt\,$, $r\opt$ depends on $\bb$. 

Up to now we have optimized the interpolating gauge field $\Delta U$ 
independent of the actual configurations in the MC simulation. 
However, it is possible to choose $\Delta U$ dependent on the pseudofermion 
$\Phi$, which is held fix during the update by $\Delta U$.
For fixed $r$ it remains the freedom to shift $\Delta U^r$ in \eqref{define}
round the torus by $\xb$
\eqa
    \Delta U^r_{\xb}([x,\mu]) \ = \ \Delta U^r([x-\xb,\mu]) \ \ \rightarrow \ \
    \tilde{\chi}^r_{\xb}(x) = \tilde{\chi}^r(x-\xb) \ .
\eqb
By this we minimize the lattice analogue of 
$<|\Phi|,|(\tilde{\chi}-c\,\omega)|>$ in \eqref{estimate}
\eqa
    \xb:\ \sum_x |\Phi(x)| \cdot |\tilde{\chi}^r(x-\xb)-\Omega(x-\xb)| = 
    \unit{min}\ ,
\eqb
$\Omega(x)$ is defined in \eqref{lattice-constant}. 

\begin{figure}[htb]
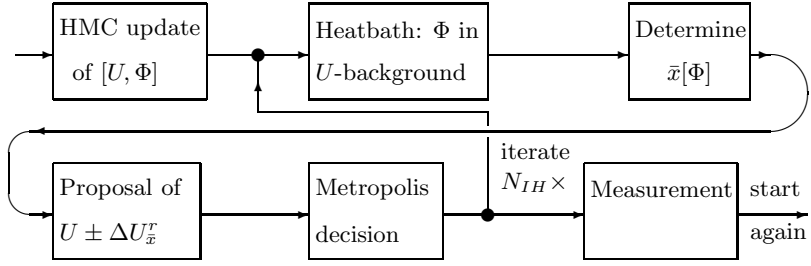

\begin{center}
\hspace{-35mm} \include{fluxpic1}
\end{center}
\vspace{-15mm}
\caption{Flux diagram for the improved instanton hits.
\label{alg-fig}}
\end{figure}
\vspace{6mm}

In our MC algorithm we implemented the new instanton hits as shown in
\fig{alg-fig}. The number $N_{\subs{IH}}$ of instanton hits (combined with a 
preceding heatbath for the pseudofermion field) per Molecular Dynamics update 
is set equal to the number of Molecular Dynamics steps, in order to obtain 
approximately the same computational effort for both parts of the algorithm.
Since the AZM $\tilde{\chi}^r$ must be calculated only once for a given
lattice, the effort to determine $\xb$ is significantly smaller than the 
CG solutions of the appearing linear equations.

\begin{figure}[b]
\begin{center}
\vspace{10mm}
\vspace{-100mm}
\vbox{ \epsfig{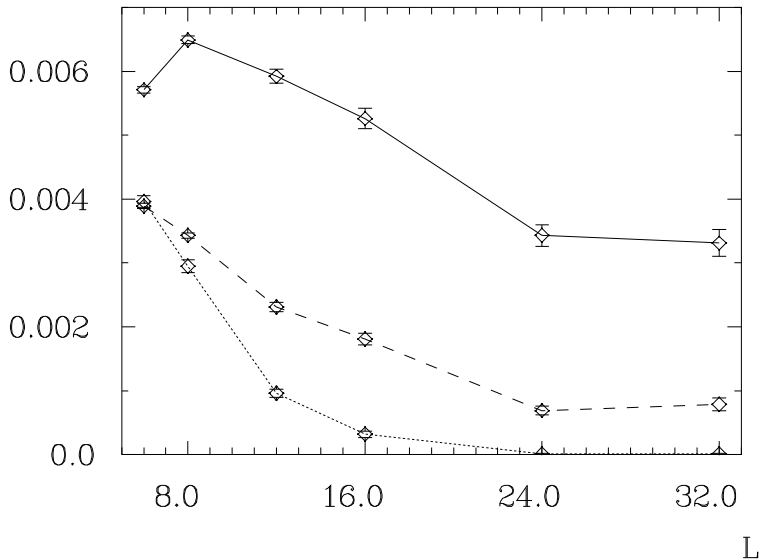} }
\vspace{-24mm}
\end{center}
\caption{The acceptance rate $P_A$ in dependence on the lattice size 
         $L\times L$: For local hits with matched $\xb$ (full line),
         with arbitrary $\xb$ (dashed), for global hits (dotted).
\label{PAvol-fig}}
\end{figure}

The use of matching $\xb$ for given $\Phi$ can be seen in \fig{PAvol-fig}.
It appears that there is no breakdown of $P_A$ for large lattices. 
The acceptance rates are still smaller than $1\%$. However, the 
computational effort for a single instanton hit, essentially given by one 
CG solution, is low compared with the effort for the HMC 
updates. They need $N_{\subs{MD}}$ CG solutions, 
$N_{\subs{MD}}$ is the number of Molecular Dynamics steps ($N_{\subs{MD}}=10$ 
to $60$ for lattice sizes $6\times6$ to $32\times32$).
In \tab{acceptance-tab} it can be seen, that there is also an improvement 
of the mean reduced acceptance rate $\Pb_R$, in which the effect of the 
zero mode is separated, see \eqref{reduced}. This means an improvement of the 
instanton hits also with respect to the higher modes.
 
\begin{table}[htb]
\begin{center} 
\caption{The weights $q_1$ and $q_2$ vs.\ the lattice size $L\times L$ for 
         $\bb=4$.
\label{weights-tab}}  
\vspace{3mm}
\begin{tabular}{c|c|c|c|c|c|c}
  \hline
  $L$ & $6$ & $8$ & $12$ & $16$ & $24$ & $32$ \\
  \hline
  $q_1$: & $0.0122(3)$ & $0.0295(6)$ & $0.073(3)$ 
         & $0.128(7) $ & $0.171(12)$ & $0.239(10)$   \str\\
  $q_2$: & $0         $ & $0.00006(3)$ & $0.0014(4)$ 
         & $0.0080(18)$ & $0.025(7)  $ & $0.062(7) $ \str\\
  \hline
\end{tabular}
\end{center}
\end{table}

\begin{table}[htb]
\begin{center}
\caption{The lowest mean eigenvalues $\bar{\lambda}_i$ 
         of $(M[U]M^\dagger[U])$.
\label{eigenvalue-tab}}
\vspace{3mm}
\begin{tabular}{c|l|l|l|l|l|l}
 \hline
  $12\times12,\ \bb=4$: 
  & $\ \ \bar{\lambda}_1$ & $\ \ \bar{\lambda}_2$ & $\ \ \bar{\lambda}_3$ 
  & $\ \ \bar{\lambda}_4$ & $\ \ \bar{\lambda}_5$ & $\ \ \bar{\lambda}_6$\str\\
  \hline
  $\Qtop=\ \ 0:$& $0.0198(4)$ & $0.0351(5)$ & $0.0627(4)$ 
                & $0.0865(5)$ & $0.129(1) $ & $0.159(1) $        \str\\
  $\Qtop=\pm1:$ & $0.0038(3)$ & $0.033(1) $ & $0.056(1) $ 
                & $0.093(1) $ & $0.124(1) $ & $0.160(2) $        \str\\
  $\Qtop=\pm2:$ & $0.0018(3)$ & $0.0051(4)$ & $0.056(7) $ 
                & $0.098(4) $ & $0.122(5) $ & $0.164(5) $        \str\\
 \hline \hline
  $18\times18,\ \bb=9$: 
  & $\ \ \bar{\lambda}_1$ & $\ \ \bar{\lambda}_2$ & $\ \ \bar{\lambda}_3$ 
  & $\ \ \bar{\lambda}_4$ & $\ \ \bar{\lambda}_5$ & $\ \ \bar{\lambda}_6$\str\\
  \hline
  $\Qtop=\ \ 0:$ & $0.0115(1)$ & $0.0170(2)$ & $0.0321(2)$ 
                 & $0.0409(2)$ & $0.0663(3)$ & $0.0783(3)$ \str\\
  $\Qtop=\pm1:$ & $0.00060(3)$ & $0.0190(5)$ & $0.0271(5)$ 
                &  $0.0488(6)$ & $0.0619(7)$ & $0.0824(8)$ \str\\
  $\Qtop=\pm2:$ & $0.00055(5)$ & $0.0016(1)$ & $0.032(1) $ 
                &  $0.045(1) $ & $0.064(1) $ & $0.080(1) $ \str\\
  \hline \hline
  $32\times32,\ \bb=4$: 
  & $\ \ \bar{\lambda}_1$ & $\ \ \bar{\lambda}_2$ & $\ \ \bar{\lambda}_3$ 
  & $\ \ \bar{\lambda}_4$ & $\ \ \bar{\lambda}_5$ & $\ \ \bar{\lambda}_6$\str\\
  $\Qtop=\ \ 0:$ & $0.0008(1)$ & $0.0022(2)$ & $0.0042(2)$ 
                 & $0.0064(2)$ & $0.0094(3)$ & $0.0122(3)$ \str\\
  $\Qtop=\pm1:$ & $0.00052(6)$ & $0.0018(1)$ & $0.0036(2)$ 
                 & $0.0055(3)$ & $0.0084(4)$ & $0.0118(4)$ \str\\
  $\Qtop=\pm2:$  & $0.0004(1)$ & $0.0013(2)$ & $0.0035(6)$ 
                 & $0.0061(7)$ & $0.009(1)$  & $0.011(1)$  \str\\
  \hline
\end{tabular}
\end{center}
\end{table}
   
We measured the weights $q_n$ of the topological sectors $\Qtop=n, n=1,2$.
For that we used the equality $q_n = q_{-n}$, due to a charge conjugation
symmetry. The results for $\bb=4$ and different lattice sizes are given 
in \tab{weights-tab}.
For large volumes the suppression of the non-trivial sectors gets weaker
significantly. As reason we assume that on the lattice the suppression
by the only approximate zero modes for a given value of $\bb$ can be overcome
by phase space factors for large volumes. The finite volume behavior,
characterized in section~1, is then achieved again by increasing $\bb$ and 
fixing the volume in physical units. 

We want to conclude with the following remarks:  
Firstly, for the change from $\Qtop=n$ to $\Qtop=n+1$, there is one 
additional new zero mode, which is no subject of the matching of the 
interpolating gauge field $\Delta U$ and therefore in general is not 
(approximately) orthogonal on $\Phi$. 
In particular, for $\Qtop[U]=0$, a zero mode $\chi\old$ determining the only 
zero mode $\chi\new$ in the background of $\Delta U \cdot U$ exists only for 
special toron values $t_\mu=0$. 
In fact, for approximately vanishing toron values, the improvement 
of the instanton hits worked better than for the general case. 
On the other hand we found the acceptance rates for hits changing 
$\Qtop = \pm 1 \rightarrow \pm 2$ to be of the same order of magnitude as those
for $\Qtop = 0 \rightarrow \pm 1$ on the larger lattices, where the higher
sectors are not so strongly suppressed.  
  
Secondly, the arguments leading to the improved instanton hits are strictly 
based on the case of combining an instanton (and its zero mode) with another 
instanton, thereby increasing the overall topological charge.  
On small lattices we didn't find any evidence of approximate zero modes  
corresponding to a pair of local instanton and anti-instanton, which does not
affect the topological charge. The number of AZMs, determined by a significant 
gap in the eigenvalues $\lambda^i$ of $(M[U]M^\dagger[U])$, is given by 
$\Qtop$, see \tab{eigenvalue-tab}. 
The situation for $\bb=4$ on a $32\times32$-lattice is less clear. For us it is
an open question, whether this is due to AZMs connected to instanton 
anti-instanton pairs or due to the ambiguity in the definition of the 
topological charge for the relatively low value $\bb=4$. Such an ambiguity 
becomes stronger on larger lattices. Note that a simple combination of 
instanton and anti-instanton solutions is not possible, due to the necessary 
spin flip in such combinations, see the appendix.
Eigenvalues for other parameters and in the quenched case can be found in 
\citalt{16}
.

Finally, a different picture would arise, if the zero modes of topologically 
non-trivial update configurations were localized, i.e.\ $c=0$ in 
\eqref{constant}.  
Old, local zero modes would remain approximately invariant under such updates.
The update configurations should then be matched in order to minimize the 
scalar product of the pseudofermion field and the new zero mode.

\mysectionwithout{Acknowledgements}

I'm thankful for helpful discussions with Hans Joos, Klaus Pinn, and 
Martin B\"aker.

\mysectionwithout{References}
{\small
\begin{enumerate}
\item             J. Schwinger, Phys.\ Rev.\ 128 (1962) 2425;\\
                  J. Lowenstein and A. Swieca, Ann.\ of Phys.\ 68 (1971) 172
\item             A.A. Belavin, A.M. Polyakov, A.S. Shvarts, and Yu.S. Tyupkin,
                  Phys.\ Lett.\ B59 (1975) 85; \\
                  G. 't Hooft, Phys.\ Rev.\ Lett.\ 37 (1976) 8
\item             S.L. Adler and W.A. Bardeen, Phys.\ Rev.\ 182 (1969) 1517\\
                  J.S. Bell and R.\ Jackiw, Nuovo Cim.\ 51 (1969) 47
\item             D. Yu. Grigoriev, V.A. Rubakov, and M.E. Shaposhnikov,
                  Phys.\ Lett.\ B 216 (1989) 172; \\
                  I. Montvay, Nucl.\ Phys. B (Proc.\ Suppl.) 34 (1994) 631;
\item             H. Dilger, PhD thesis, Hamburg 1993, DESY-93-181;
                  Phys.\ Lett.\ B 294 (1992) 263; `Chiral symmetry breaking on 
                  the lattice - a study in the two-flavor Schwinger model', 
                  to be published in Nucl.\ Phys.\ B
\item             S.\ Duane, A.D. Kennedy, B.J. Pendleton, and D. Roweth, 
                  Phys.\ Lett.\ B\ 195 (1987) 216
\item             H.\ Joos, Helv.\ Phys.\ Acta 63 (1990) 670;
                  Nucl.\ Phys.\ B\ (Proc.\ Suppl.)\ 17 (1990) 704
\item             S.I.\ Azakov and H.\ Joos, `The Schwinger Model on the 
                  Torus II', DESY-94-142, submitted to Helv.\ Phys.\ Acta
\item             F. Fucito and S. Solomon, Phys.\ Lett.\ B 314 (1984) 230;\\
                  M.L. Laursen, J. Smit, and J.C. Vink, Phys.\ Lett.\ B 262 
                                                                   (1991) 467
\item             M.\ L\"uscher, Comm.\ Math.\ Phys.\ 85 (1982) 39;\\
                  A. Phillips and D. Stone, Comm.\ Math.\ Phys.\ 103 (1986) 599
\item             J. Smit and J.C.\ Vink, Nucl.\ Phys.\ B\ 286 (1987) 485;
                  Nucl.\ Phys.\ B 303 (1988) 36
\item             C. Jayewardena, Helv.\ Phys.\ Acta 61 (1988) 636
\item             U.J. Wiese, Nucl.\ Phys.\ B 318 (1989) 153
\item             I. Polonyi, H.W. Wyld, and J.B. Kogut, Phys.\ Rev.\
                  Lett.\ 53 (1984) 644; \\
                  O. Martin and S.W. Otto, Phys.\ Rev.\ D31 (1985) 435
\item             I. Montvay, DESY-93-184; Phys.\ Lett.\ B 323 (1994) 378
\item             M. B\"aker, to appear
\item             E. K\"ahler, Rend.\ Mat.\ Ser.\ V, 21 (1962) 425
\item             P. Becher and H. Joos, Z.\ Phys.\ C15 (1982) 343
\item             M. Atiyah and I. Singer, Ann.\ Math.\ 87 (1968) 484
\item             A. Erd\'{e}lyi et al., Higher Transcendental Functions,
                  Vol.\ 2, Chapter 13 (McGraw-Hill, 1953); \\
                  D. Mumford, Tata Lectures on Theta (Birkh\"auser 1983)

\end{enumerate} }

\mysectionwithout{Appendix}
\setcounter{equation}{0}
\renewcommand{\theequation}{A.\arabic{equation}}

The two-flavor SM in the continuum on the torus $\T$ was studied 
by S.I. Azakov, H. Joos.\cit{7,8} 
They partly use the geometric formulation with Dirac-K\"ahler (DK) 
fermions.\cit{17} 
This means the fermions are described by inhomogenous differential forms 
$\phi(x,H)dx^H$, $H$$=$$\emptyset,1,2,12$. In two dimensions these can be 
transformed into two flavors of Dirac fermions $\psi_{ab}(x)$, $a=1,2$ is the 
spinor index, $b=1,2$ is the flavor index. Since there is a systematic lattice 
restriction of DK fermions to staggered fermions,\cit{18} 
the geometric formulation is useful for a comparison of continuum and lattice 
model, in our case for the comparison of continuum zero modes with approximate 
zero modes on the lattice.
In this brief description of the continuum zero mode structure we will use the 
Dirac language as far as possible.

We consider the zero modes $\chi_{ab}(x)$ of the Dirac operator in the
gauge field $A_\mu(x)$
\eqa \label{zero-mode}
    \gmu_{aa'} \, [\dmu - ieA_\mu(x)] \, \chi_{a'b}(x) \ = \ 0 \ .    
\eqb
The $U(1)$ gauge field on $\T$ can be written
\eqa \label{Hodge}
    A_\mu(x) \ = \ \dmu a(x) \ + \ \epsilon_\mu^{\ \nu} \partial_\nu b(x  )
             \ + \ t_\mu
             \ + \frac{2 \pi k}{e V} \epsilon_\mu^{\ \nu} x_\nu,
\eqb
$\dmu a(x)$ is a pure gauge,
$\epsilon_\mu^{\ \nu} \partial_\nu b(x)$ is a gauge field in Lorentz
gauge, $t_\mu$ is the so-called toron field.
The last term $\propto \epsilon_\mu^{\ \nu} x_\nu$ is a representative of the 
Chern class with Chern index $k\in\Z$. It is periodic up to (non-trivial)
gauge transformations, and it has a constant field strength, i.e.\ it is a 
topologically non-trivial classical solution of the free field equations. 
For $k=1$ it is called an instanton.

According to the Atiyah-Singer index theorem for two flavors of 
fermions\cit{19} 
\eqa
    n_l \ - \ n_r \ = \ 2k \ ,                                 \label{index}
\eqb
the Dirac operator in such a background field has $n_l=2k$ left-handed 
zero modes for $k>0$, $n_r=2k$ right-handed zero modes for $k<0$. 
For $k=0$ there are $4$ vectorlike zero modes for vanishing toron field, 
$t_\mu=0$.
For $k>0$ the zero modes have the form
\eqa  \label{explicit}
    \chi^{e/o}_{ab}(x) \ = \ f(x) \, \omega^{e/o}_{ab} \ = \
    c \, e^{ie(a(x) + ib(x) + t_\mu x^\mu / 2)} \, 
    e^{(k\pi/2\tau) (z^2 - |z|^2)} \, H(z) \,
    \, \omega^{e/o}_{ab} \ ,
\eqb
with $z = (x_1+ix_2)/L_1 -  (t_2 - it_1)eL_2/(2\pi k)$. 
$H(z)$ denotes a polynomial in Jacobi's $\theta_n(z) \equiv \theta_n(z|\tau)
$-functions, $\tau =iL_2/L_1$, see \citalt{20}
. For $k=n$ one has
$n$ linear independent polynomials $H^{(i,n)}(z)$, see \tab{thetapol-tab}. 
The spin-flavor dependence of these zero modes is simply contained in the
constant spin-flavor vectors $\omega^{e/o}_{ab}$. They are chosen as 
eigenvectors of a chiral isospin transformation $\Ac$ (the main automorphism 
for differential forms in the geometric formulation), denoted by even and odd.
With $\gf$ chosen diagonal one gets 
\eqa  \label{omega-Dirac}
    \omega^{e}_{ab} = \delta_{a,1} \, \delta_{b,1} \, , \  
    \omega^{o}_{ab} = \delta_{a,1} \, \delta_{b,2} \, , \ \
    \Ac \omega^{e/o}_{ab} = \gf_{aa'} \omega^{e/o}_{a'b'} \gf_{b'b} 
    = \pm\omega^{e/o}_{ab} \ .
\eqb
For $k<0$ one has to take the charge conjugated solutions $\chi^*_{ab}$ of 
\eqref{explicit}.
For $k=0$, $t_\mu=0$ the zero modes still fit the form of \eqref{explicit} with
$H(z)=1$. However, the charge conjugated functions are then solutions at the 
same time, thus there are 4 zero modes in this case. 

\begin{table}
\begin{center}
\caption{The polynomials in $\theta$-functions $H^{(i,n)}, i=1,\dots,n$, 
         for $n=1,\dots,5$.
\label{thetapol-tab}}
\vspace{3mm}
\begin{tabular}{c|c|c|c|c|c}
  \hline
  $k=1:$ & & & $\theta_3$ & &                                   \str\\    
  \hline
  $k=2:$ & & $\theta_1^2$ & $\theta_3^2$ & &                    \str\\    
  \hline
  $k=3:$ & & $\theta_1^2\,\theta_3$ & $\theta_3^3$ 
         & $\theta_1\,\theta_2\,\theta_4$ &                     \str\\    
  \hline
  $k=4:$ & $\theta_1^4$ & $\theta_1^2\,\theta_3^2$ & $\theta_3^4$ 
         & $\theta_1\,\theta_3\,\theta_2\,\theta_4$ &           \str\\    
  \hline
  $k=5:$ & $\theta_1^4\,\theta_3$ & $\theta_1^2\,\theta_3^3$ & $\theta_3^5$ 
         & $\theta_1\,\theta_3^2\,\theta_2\,\theta_4$ 
         & $\theta_1^3\,\theta_2\,\theta_4$                     \str\\    
  \hline
\end{tabular}
\end{center}
\vspace{-5mm}
\end{table}
  
The geometric version of these zero modes is given by
\eqa  \label{omega-DK}
    \chi^{e/o}(x,H) \, dx^H = f(x) \, \omega^{e/o}_H \, ,  \ \
    \omega^{e}_H \ = \ 1 - idx^{12} \, , \ \omega^{o}_H \ = dx^1 + idx^2 \ .
\eqb
In this form they can be easily compared with AZMs on the 
lattice. In the discretization of DK fermions to staggered fermions, 
the homogenous forms $\propto dx^H$ correspond to the staggered fermion field
on the lattice sites with $x_\mu = \unit{odd}$ for $\mu \in H$. Since, the
chiral isospin transformations generated by $\Ac$ in \eqref{omega-Dirac} hold 
on the lattice, the lattice AZMs can be distinguished in even and odd modes 
too. The even AZMs then live only on the even lattice sites, $x_1+x_2 = 
\unit{even}$. 
In regions of approximately constant $f(x)$ they should be proportional to
$\Omega(x)$
\eqa \label{omega}
    \Omega(x) = a^0    \ \mbox{ for } x_1,x_2 \ \unit{even} \ , \ \
    \Omega(x) = a^{12} = -ia^0 \ \mbox{ for } x_1,x_2 \ \unit{odd} \ .
\eqb

Due to the particular spin-flavor dependence, the product rule of 
differentiation leads to a multiplicative structure of the zero modes for 
$k\geq0$ ($k\leq0)$, proportional to the same spin-flavor vector $\omega$. 
Let $\chi_A(x)=f_A(x)\,\omega$ be a zero mode in the background field
$A_\mu(x)$. Then $\chi_A \cdot \chi_B\,(x) := f_A(x) f_B(x)\,\omega$ is a
zero mode in the background $A_\mu(x) + B_\mu(x)$ 
\eqa 
          \gmu \, [\dmu - ie(A_\mu + B_\mu)] \ \chi_A \cdot \chi_B  
    \ = \ f_B \ \gmu \, [\dmu - ieA_\mu] \ \chi_A 
    \ + \ f_A \ \gmu \, [\dmu - ieB_\mu] \ \chi_B
    \ = \ 0 \ .                                                 \label{multi}
\eqb
For vanishing toron field $t_\mu=0$ this multiplicative structure is evident
from the explicit form of the zero modes, see \eqref{explicit} and 
\tab{thetapol-tab}. For other values of $t_\mu$ this structure is realized with
the addition formulas for $\theta$-functions.\cit{20} 

A similar picture doesn't arise for a zero mode in the background 
$A_\mu(x) - B_\mu(x)$, where $k\geq0$ ($k\leq0)$ for both fields. Functions
like $f_A(x)/f_B(x)$ give rise to a zero mode, if and only if they are 
continuous, which in general is not the case.
Corresponding to that for $k[A]>0, k[B]<0$ the corresponding spin-flavor
vectors are complex conjugate to each other, and \eqref{multi} is not valid. 

Finally we want to describe what happens, if a topologically non-trivial gauge 
field on $\T$ is embedded in a larger volume $\T'$. Even if the gauge field on 
$\T'$ vanishes on most of this bigger torus, the zero mode does not become
arbitrarily small for large distances to the region of non-vanishing gauge 
field. However, on the lattice we could numerically achieve a behavior as
given in \eqref{constant}, see \fig{chitil-fig}.  
As a consequence of this non-localization of zero modes even in localized gauge
field configurations, it is not clear how to construct (approximate) zero modes
of instanton anti-instanton pairs.
There is no simple way to glue the zero modes $\chi$ and $\chi^*$ of a local 
instanton ($k=1$) and a local anti-instanton ($k=-1$) together, even if they 
are situated far from each other. 
In a transition region the spin-flavor given by $\omega$ must be fliped to 
$\omega^*$. It is not clear, whether such a spin-flip is possible without a 
significant rise of the eigenvalue.   

\clearpage
\end{document}

%% file: fluxpic1.tex
{\footnotesize 
\unitlength=0.85mm
\linethickness{0.4pt}
\begin{picture}(144.00,50.00)
\put(15.00,35.00){\line(1,0){24.00}}
\put(39.00,35.00){\line(0,1){15.00}}
\put(39.00,50.00){\line(-1,0){24.00}}
\put(15.00,35.00){\line(0,1){15.00}}
\put(16.00,46.00){\makebox(0,0)[lc]{HMC update}}
\put(18.00,39.00){\makebox(0,0)[lc]{of $[U,\Phi]$}}
\put(55.00,35.00){\line(1,0){28.00}}
\put(83.00,35.00){\line(0,1){15.00}}
\put(83.00,50.00){\line(-1,0){28.00}}
\put(55.00,35.00){\line(0,1){15.00}}
\put(56.00,46.00){\makebox(0,0)[lc]{Heatbath: $\Phi$ in}}
\put(56.00,39.00){\makebox(0,0)[lc]{$U$-background}}
\put(105.00,35.00){\line(1,0){19.00}}
\put(124.00,35.00){\line(0,1){15.00}}
\put(124.00,50.00){\line(-1,0){19.00}}
\put(105.00,35.00){\line(0,1){15.00}}
\put(106.00,46.00){\makebox(0,0)[lc]{Determine}}
\put(111.00,39.00){\makebox(0,0)[lc]{$\bar{x}[\Phi]$}}
\put(39.00,42.00){\vector(1,0){16.00}}
\put(83.00,42.00){\vector(1,0){22.00}}
\put(127.00,36.00){\oval(12.00,12.00)[r]}
\put(124.00,42.00){\vector(1,0){3.00}}
\put(127.00,30.00){\vector(-1,0){115.00}}
\put(15.00,10.00){\line(0,1){15.00}}
\put(15.00,25.00){\line(1,0){23.00}}
\put(38.00,25.00){\line(0,-1){15.00}}
\put(38.00,10.00){\line(-1,0){23.00}}
\put(12.00,23.50){\oval(8.00,13.00)[l]}
\put(16.00,21.00){\makebox(0,0)[lc]{Proposal of}}
\put(16.00,14.00){\makebox(0,0)[lc]{$U\pm\Delta U^r_{\bar{x}}$}}
\put(55.00,10.00){\line(0,1){15.00}}
\put(56.00,21.00){\makebox(0,0)[lc]{Metropolis}}
\put(56.00,14.00){\makebox(0,0)[lc]{decision}}
\put(122.00,10.00){\line(0,1){15.00}}
\put(38.00,17.00){\vector(1,0){17.00}}
\put(12.00,17.00){\vector(1,0){3.00}}
\put(09.00,42.00){\vector(1,0){6.00}}
\put(83.00,17.00){\line(0,1){12.00}}
\put(47.00,38.00){\line(0,1){4.00}}
\put(47.00,42.00){\circle*{2.00}}
\put(83.00,17.00){\circle*{2.00}}
\put(122.00,17.00){\vector(1,0){11.00}}
\put(124.00,21.00){\makebox(0,0)[lc]{start}}
\put(124.00,14.00){\makebox(0,0)[lc]{again}}
\put(85.00,27.00){\makebox(0,0)[lc]{iterate}}
\put(85.00,22.00){\makebox(0,0)[lc]{$N_{IH} \times$}}
\put(55.00,25.00){\line(1,0){21.00}}
\put(76.00,25.00){\line(0,-1){15.00}}
\put(76.00,10.00){\line(-1,0){21.00}}
\put(76.00,17.00){\line(1,0){7.00}}
\put(83.00,31.00){\line(0,1){2.00}}
\put(83.00,33.00){\line(-1,0){36.00}}
\put(47.00,33.00){\vector(0,1){5.00}}
\put(98.00,10.00){\line(0,1){15.00}}
\put(98.00,25.00){\line(1,0){24.00}}
\put(98.00,10.00){\line(1,0){24.00}}
\put(99.00,21.00){\makebox(0,0)[lc]{Measurement}}
\put(83.00,17.00){\vector(1,0){15.00}}
\end{picture}
}